# Electromagnetic, atomic-structure and chemistry changes induced by Ca-doping of low-angle YBa$_2$Cu$_3$O$_{7-\delta}$ grain boundaries

XUEYAN SONG, GEORGE DANIELS, D. MATT FELDMANN, ALEX GUREVICH, DAVID LARBALESTIER∗

Applied Superconductivity Center, University of Wisconsin, Madison, WI 53706, USA

Practical high temperature superconductors must be textured to minimize the reduction of the critical current density J$_{gb}$ at misoriented grain boundaries. Partial substitution of Ca for Y in YBa$_2$Cu$_3$O$_{7-\delta}$ has shown significant improvement in J$_{gb}$ but the mechanisms are still not well understood. Here we report atomic-scale, structural and analytical electron microscopy combined with transport measurements on 7° [001]-tilt Y$_{0.7}$Ca$_{0.3}$Ba$_2$Cu$_3$O$_{7-\delta}$ and YBa$_2$Cu$_3$O$_{7-\delta}$ grain boundaries, where the dislocation cores are well separated. We show that the enhanced carrier density, higher J$_{gb}$ and weaker superconductivity depression at the Ca-doped boundary result from a strong, non-monotonic Ca segregation and structural rearrangements on a scale of ~1 nm near the dislocation cores. We propose a model of the formation of Ca$^{2+}$ solute atmospheres in the strain and electric fields of the grain boundary and show that Ca doping expands the dislocation cores yet enhances J$_{gb}$ by improving the screening and local hole concentration.

Second generation superconducting coated conductors (CCs) made from $YBa_2Cu_3O_{7-\delta}$ (YBCO), have great application potential[1] but conductor design remains severely constrained by the need for biaxial texture to avoid the exponential decrease of the critical current density $J_{gb}(\theta)$ through grain boundaries (GBs) with increasing misorientation angle $\theta$ above 3-5°[2,3,4]. The most viable routes to CCs require deposition of $(RE)Ba_2Cu_3O_{7-\delta}$ onto metal tapes textured by ion-beam assisted deposition[5,6] or oxide-buffered, rolling-assisted bi-axially textured Ni-alloy substrates[7]. However, even the best CCs have many GBs with $\theta \approx$ 5-7° which partially obstruct current. The current-blocking effect of GBs[8-13] in YBCO can indeed be mitigated by Ca doping, but the mechanisms by which Ca improves $J_{gb}$ are still not well understood. The positive effect of Ca-doping is believed to result from a partial replacement of $Y^{3+}$ by $Ca^{2+}$, which increases the hole carrier density at the GB[13]. Thus, despite reduction of $T_c$ in Ca doped YBCO[13], $J_{gb}$ at $T << T_c$ can be ameliorated both by Ca addition[8,10] and by annealing in oxygen[9]. It has been recently realized that $J_{gb}$ is mostly controlled by weakened superconductivity due to strain and charging effects[10,14] in the nanoscale channels between the GB dislocations. Significant GB charging was indeed revealed by electron holography of 4° [001]-tilt pure and Ca-doped YBCO bicrystals[15] for which the negative GB potential was smaller (-1 V vs. -2.4 V) and decayed over a shorter length (0.8 nm vs. 1.7 nm) for the Ca-doped GB. However, detailed nanoscale structural characterization of doped and undoped grain boundaries that explains the effect of Ca substitution is still lacking.

In this paper we investigate the lattice structure, electronic state and current transport of pure and Ca-doped GBs in low-angle YBCO bicrystals by combining measurements of extended voltage-current (V-I) characteristics on a matched pair of Ca-doped and pure 7° [001]-tilt YBCO bicrystals, and detailed atomic-scale microscopy and local composition of the same GBs. We found a significant Ca segregation to the GB and a non-monotonic variation of Ca concentration around the dislocation cores on scales ~ 1 nm. The structural changes induced by Ca doping expand the dislocation cores both normal and parallel to the GB, yet surprisingly *increase* $J_{gb}$ close to the critical current density $J_g$ in the grains. The totality of our high resolution electron microscopy and transport data indicates that this local Ca segregation significantly reduces the superconductivity depression at the GB and improves vortex pinning in magnetic fields.

V-I characteristics, $J_g$ and $J_{gb}$ measured after post-annealing in oxygen for 30 minutes at 420°C are shown in Figure 1. This treatment put both samples into the overdoped state with $T_c$ = 91 K for YBCO and 71 K for the 0.3Ca-doped samples. The data are compared at the same reduced temperature $T/T_c$ = 0.86 (78 K for the pure and 61 K for the Ca-doped YBCO). The $J_{gb}/J_g$ ratio measured at the electric field E = 1 µV/cm varies from $J_{gb}/J_g$ < 1 at low magnetic fields B to $J_{gb}/J_g \approx$ 1 at B $\approx$ 3-4 T for both samples. $J_{gb}/J_g$ is much smaller for the pure YBCO than for the Ca-doped film, and $J_{gb}$ for the Ca-doped film is higher, even though it has a lower $J_g$ than YBCO. At lower B, YBCO exhibits a kink in the log E-log J plot at $J \approx J_{gb}$, unlike the smooth log E-log J intragrain traces. By contrast, the Ca-doped GB exhibits much more gradual log E–log J characteristics. These differences are evident in the linear E-J plots, which show a pronounced ohmic resistance $R_F$ for pure YBCO, but rather nonlinear E(J) for the Ca-doped YBCO at $J > J_{gb}$. In either case $J_{gb}$ falls below $J_g$ because of depressed superconductivity at the GB.

We performed extensive transmission electron microscopy (TEM) of the GBs from the micron to the atomic-scale. Both the pure and Ca-doped GBs were rather straight with ~85% (100) GB segments, connected by short facets along (110) and (100), as shown in Figure 2. The [100] dislocations separate channels through which current flows. The GB shows up very well under high-angle annular dark-field (HAADF) Z-contrast imaging whose intensity W $\propto Z^{1.7}$ depends on the atomic number $Z^{17,18}$, differentiating the brighter Y/Ba from darker Cu/O(4) columns. The Z-contrast images and atomic stacking patterns of Figure 3a show that [100] dislocation cores in pure YBCO GB have 2 Cu/O columns surrounded by 5 Y/Ba columns. Contrast immediately below the terminating Y/Ba column is poor but intensity traces taken a little lower indicate that the Cu/O columns are continuous across the expanded, tensile region of the dislocation.

Figure 3b shows a Z-contrast image of the Ca-doped film in which the brighter columns contain Y/Ba/Ca atoms. The [100] dislocation is again defined by one missing Y/Ba/Ca column, but the contrast immediately below this terminating column in the tensile region is noticeably different. Two bright columns with similar intensity to the neighboring intragrain Y/Ba/Ca columns appear at the expected Cu/O column sites. The arrow in Figure 3b marks a third anomalous column which is *darker* than the neighboring Cu/O columns. These two bright and one dark column are consistently observed in every tensile segment of the Ca-doped GB. These images show that the dislocation core region is also significantly expanded perpendicular to the GB plane, as shown in Figure 3c, while the tensile part of the dislocation extends along the Ca-doped GB over 3 unit cells, rather than the 1-2 cells in pure YBCO.

Electron energy loss spectra (EELS) were measured both perpendicular and parallel to the GB with a 0.2 nm electron probe and a stage-drift-limited spatial resolution of 0.3-0.4 nm. Figure 4a compares Ca $L_{23}$ edge spectra averaged from 10 different individual dislocation cores and 15 intragrain regions. The data show that the Ca content of 0.46 in the tensile part of the dislocation core exceeds the nominal bulk Ca concentration by 55%. Spectral traces normal to the GB revealed sharp Ca peaks at the GB, and pronounced minima about 2 unit cells away, as shown in Figure 4b. Moreover, the Ca concentration varies *non-monotonically* along the GB near the 2 bright-1 dark column triplet at the dislocation core in Figure 4c. The single dark column in the Z-contrast images is another indication of strong local Ca-enrichment.

Oxygen K-edge EELS spectra at the dislocation cores in Figure 5 clearly show a pre-peak around 528 eV indicative of metallic behavior, both for the pure and Ca-doped GB. Although EELS imaging damaged the cores within a few seconds, we were able to measure a lower hole deficiency at the Ca-doped GB. For a nearly optimally doped $YBa_2Cu_3O_{7-\delta}$ with $\delta \approx 0.2$, the oxygen K-edge at 535 eV is coupled to the pre-peak at 528 eV by transitions to the O-2p states, which form the hole valence band. As $\delta$ increases, the intensity of the 528 eV peak falls, indicating a decrease in the hole concentration and a reduction of $T_c$[19]. Electron beam damage makes spatial variations of the oxygen K-edge pre-peak intensity inconclusive, yet the EELS results also indicate that neither the pure nor the Ca-doped GB dislocation cores are strongly hole-deficient. This result is consistent with earlier lower spatial resolution (~ 2-3 nm) data taken on low-angle, flux-grown, bulk [001]-tilt bicrystals, which exhibited only weak depression of the oxygen pre-peak near the GB[20].

The structure of the pure dislocation core in Figure 3a is also consistent with earlier studies of [100] dislocations at YBCO GBs[21]. However, we found for the first time that Ca not only expands the dislocation cores, but that the structure of the Ca-doped core is different from the pure one, as shown by the unusual column intensity sequence in Figure 3b. For instance, the Ca-induced core expansion may result from a stacking fault (Y/Ba/Ca)/(Cu/O)/(Y/Ba/Ca)/(Y/Ba/Ca)/(Y/Ba/Ca)/(Cu/O)/(Y/Ba/Ca) in the dislocation tensile region. The stacking fault could result from the dislocation reaction, [100] → ½ [100] + ½ [100]. In any case, the striking aspect of the Ca segregation shown in Figure 4 is its strong variation along the GB on the scale of the unit cell, indicating that the larger $Ca^{2+}$ ion of radius $r_{ca}$ = 0.099 nm may preferentially substitute for the smaller $Y^{3+}$ ($r_Y$ = 0.09nm) in the tensile part of the dislocation core[22]. The segregation of solute can strongly affect the structure of GBs[23-25], for example, Ca segregation can induce the GB lattice transformation in MgO[25].

The surprising result of our microscopy is that the undisturbed channel between the dislocation cores was narrowed, yet our transport data show a higher $J_{gb}$ for the Ca-doped bicrystals over a wide range of field and temperature, in agreement with prior studies[9,11,15]. Here $J_{gb}$ is limited by pinning of the GB vortices whose structure changes dramatically as compared to the bulk Abrikosov vortices. Indeed, depressed superconductivity at the GB due to charging, d-wave symmetry and local non-stoichiometry causes expansion of the GB vortex core from a normal core of the order of the intragrain coherence length $\xi \approx$ 4 nm to a stretched Josephson core of length $\ell(T) \gg \xi$ along the GB[16]. The core size $\ell(T)$ along the GB is determined by the averaged GB depairing current density $J_0$ which can be extracted from the flux flow resistance, $R_F = [H/(H + H_0)]^{1/2}R$ at $J > J_{gb}$, where R is the normal-state GB resistance, $H_0 = \phi_0/(2\pi\ell)^2$ is the crossover field at which the GB vortex cores overlap, $\ell \approx \xi J_d/J_0$, $J_d$ is the intragrain depairing current density, and $\phi_0$ is the flux quantum[16]. Previous analysis of $R_F(H)$ of the pure 7° YBCO bicrystal with the GB dislocation spacing of 3.2 nm showed that $J_0 \approx 0.1 J_d$ and $\ell \approx 10\xi$ at 77 K, while the temperature dependence of $J_0 \propto (1 - T/T_c)^2$ indicates that current channels behave like strongly proximity-coupled, superconducting-normal-superconducting point contacts. However, Figure 1 shows that Ca-doping reduces the difference between the intragrain and intergrain E-J characteristics, shrinking the GB vortex core size $\ell$ and increasing $J_0$. In fact, the difference between $J_0$ and $J_d$ for the Ca-doped bicrystal becomes so small that the analysis of Ref. 16 cannot distinguish GB vortices from bulk vortices, indicating only weak superconductivity depression on the GB. This increase of $J_0$ results from reduction of the average sheet GB charge density Q, the local electric potential $\varphi_0 = 2\pi l_D Q/\varepsilon$, and the Thomas-Fermi screening length $l_D$ produced by partial substitution of $Y^{3+}$ by $Ca^{2+}$ where $\varepsilon$ is the lattice dielectric constant. Because the negative potential $\varphi_0$ drives the superconducting state at the hole-deficient GB toward the antiferromagnetic insulator characteristic of the high-$T_c$ cuprates[14], Ca-doping improves $J_{gb}$ both by reducing Q and $l_D$, as recently shown by electron holography of Ca-doped YBCO bicrystals[15]. The strain-induced GB charge[14] $Q \propto b^2$ can also be reduced by as much as half by splitting the GB dislocation into partial dislocations with Burgers vectors b/2.

To understand the non-monotonic Ca distribution, we consider the equilibrium concentration c(**r**) of Ca in the pressure field p(x,y) and the screened electric potential $\varphi(\mathbf{r})$ produced by the GB dislocations[14,26,27,28].

$$c = \frac{c_0}{c_0 + (1-c_0)\exp[(p\Delta V + \varphi \Delta Z)/k_B T]} \quad (1)$$

Here $\Delta V = V_{Ca} - V_s$ and $\Delta Z = Z_{Ca} - Z_s$ are the differences of ionic volumes and charges of the Ca and the substituted atom (Y), $c_0$ is the bulk concentration, p(x,y) = $p_0\sin(2\pi y/d)/[\cosh(2\pi\rho/d) - \cos(2\pi y/d)]$, $d = b/2\sin(\theta/2)$[28], $p_0 = \mu\sin(\theta/2)/(1-\nu)$, $\rho = (r_0^2 + x^2)^{1/2}$, the cutoff radius $r_0 \sim b$ accounts for the plastically deformed dislocation core, $\mu$ is the shear modulus, $\nu$ is the Poisson coefficient, and x and y are coordinates across and along the GB, respectively. The distribution of Ca concentration, c(x,y) plotted in Figure 6 exhibits peaks in tensile regions followed by pronounced dips in compressed regions along the GB, and a non-monotonic dependence of c(x) across the GB, in good agreement with our EELS data in Figure 4. Equation (1) also shows that the periodic stress p(y) tends to increase the average Ca concentration at the GB, $\langle c \rangle = \int^L c(x,y)dy/L$ where L >> d. For $p_0\Delta V/k_B T \ll 1$, $\varphi = 0$, and $c_0 \ll 1$, we obtain $\langle c(x,y)\rangle = c_0\langle\exp[-\Delta V p(x,y)/k_B T]\rangle \approx c_0(1 + \Delta V^2\langle p^2\rangle/2k_B^2 T^2) > c_0$, where $\langle p \rangle = 0$, and $\langle p^2 \rangle = 2p_0^2/[\exp(4\pi\rho/d)$

– 1]. Taking $\nu = 0.3$, $\mu = 40$ GPa, $\theta = 7°$, $\Delta V = V_{Ca} - V_Y = 4\pi(r_{Ca}^3 - r_Y^3)/3$, ionic radii $r_{Ca} = 0.099$ nm, $r_Y = 0.09$ nm and $T = 300$ K, we estimate $p_0 \Delta V / k_B T_i \approx 0.85$. Thus, the GB dislocation strains favor segregation of Ca, partially offsetting the electrostatic barrier caused by the negative $\varphi(x)$ of a hole-deficient GB. For $Z_{Ca} < Z_Y$, the competition between strain and electric field effects causes minima in $c(x)$, as shown in Figures 4 and 6. Similar minima in $\varphi(x)$ were revealed by electron holography of a 4° Ca-doped YBCO bicrystal[15].

This model can explain our observation that Ca doping actually increases $J_{gb}(H)$ despite shrinkage of the current channels and expansion of the dislocation cores. As follows from Figure 6, this structural expansion naturally results from increased lattice disorder produced by the Cottrell atmosphere[28] of segregated $Ca^{2+}$ ions. However, because $Ca^{2+}$ brings extra holes to the GB, superconductivity in the cores is enhanced by partial recovery of optimum hole doping. This enhancement occurs highly non-uniformly: Ca segregates mostly in the tensile regions of the GB, while compressed regions have markedly less Ca than the bulk. The Ca-enriched charged dislocation cores are coupled across the channels by the electric potential $\varphi(x,y)$ due to hole redistribution, which reduces $\varphi_0$, $Q$ and $l_D$. As a result, the overall $J_{gb}$ increases, because the compressed Ca-deficient current channels exhibit much less $T_c$ depression than bulk Ca-doped YBCO, yet they have less hole depletion than in pure YBCO due to the lower charge and shorter $l_D$ near the dislocation cores.

Our model suggests new possibilities for manipulating the properties of doped GBs. An equilibrium Ca distribution initially forms at the film growth or annealing temperature, but then evolves according to Eqn. (1) as $T(t)$ decreases. The equilibration time $t_d \sim l_0^2/D$ of the Cottrell atmosphere is controlled by the Ca diffusivity $D = D_0 \exp(-U/k_B T)$ with $U \sim 1-2$ eV[29] over a small strained region of radius $l_0 \sim d/2\pi \sim 1$ nm around the dislocation cores. Taking $D$ (300 K) $\sim 10^{-16}$ cm$^2$/s in the ab-plane[29] we obtain $t_d \sim 10^2$ s at room temperature, but because $t_d(T)$ increases very rapidly as T decreases below 300 K, the segregation of Ca effectively freezes at a temperature $T_i$ well above $T_c$. Thus, aging effects may occur below 300 K, consistent with sometimes variable behavior observed on Ca-doped GBs. This model suggests that Ca segregation can also be controlled by the oxygen annealing conditions[9], which can cause charge-driven GB overdoping, reduce the electrostatic barrier $\varphi_0$, and facilitate further segregation of Ca at the GB. Such strain and charge-driven diffusion segregation of Ca to the GB[29] probably underlies the positive overdoping effects observed in YBCO/(Y,Ca)BCO multilayers[8]. In a recent preprint R. Klie et al. reported atomic-resolution TEM, EELS, and density functional calculations, which indicate that strains produced by the mismatch of Ca-O and Cu-O atomic bond lengths are very important for Ca segregation on YBCO grain boundaries[31].

In conclusion, our atomic-scale electron microscopy and transport measurements show that the enhanced critical current densities of Ca-doped YBCO bicrystals result from inhomogeneous Ca segregation to the grain boundary, structural expansion of the dislocation cores and non-monotonic Ca distribution both along and across the boundary on a scale ~ 1 nm. This Ca segregation causes shrinkage of the GB vortex cores and improved flux pinning, increased hole concentration and reduction of the GB charge and screening length. The beneficial effect of Ca doping at $T < T_c$ is explained by formation of quenched distributions of $Ca^{2+}$ dopants in the strain and electric fields of the dislocation cores. Our results suggest significant new opportunities for nanoscale impurity segregation engineering.

**METHODS**

Our c-axis oriented $YBa_2Cu_3O_{7-\delta}$ and $Y_{0.7}Ca_{0.3}Ba_2Cu_3O_{7-\delta}$ films of thickness 230 nm were grown by pulsed laser deposition on symmetric 7° [001]-tilt $SrTiO_3$ bicrystals[9]. After lithographic patterning transport properties of the films were measured from 10 to 90K in fields up to 12 T parallel to the c-axis[9]. Samples for plan-view TEM were prepared by mechanical polishing and ion milling. Diffraction contrast and high resolution TEM investigations were performed using a microscope (Philips CM200) operated at 200 KV with point resolution of 0.19 nm. HAADF imaging[17] in a scanning TEM was performed using a 200 KV (JEOL 2010F) microscope focused to a ~0.14 nm diameter probe[30] with the inner cut-off angle of the HAADF detector > 52 mrad.

Transmitted electrons were analyzed by an on-axis electron-energy-loss spectrometer with an energy dispersion of 0.3 eV per channel and an energy resolution of ~1.2 eV.

## Acknowledgements:

We are grateful to James Buban and Nigel Browning, formerly of the University of Illinois at Chicago, for experimental help, discussions and for access to the JEOL2010F there. We thank Steve Pennycook from ORNL for discussions. The work was supported by AFOSR under grant F49620-03-01-0429.

Correspondence and requests for materials should be addressed to D. L. at: larbalestier@engr.wisc.edu.

## Figure Captions:

Figure 1. Critical current density vs. magnetic field, and electric field-current density (E-J) characteristics of the pure and 0.3Ca-doped films at the same reduced temperature $T/T_c$ = 0.86. $J_g$ and $J_{gb}$ are evaluated at 1 μV/cm, using V-I curves measured on 25 μm wide bridges produced by photolithographically patterning the links. The E-J characteristics for the pure YBCO GB segment show a steep rise in all magnetic fields below the irreversibility field of about 5 T due to preferential Abrikosov-Josephson vortex flow at the GB. The E-J characteristics of the Ca-doped GB are very much more similar to the intragrain E-J characteristics than is the case for the pure YBCO. The ohmic behavior with the resistance $R_F$ due to vortex flow at the pure YBCO GB at $J > J_{gb}$ is indicated in the inset.

Figure 2. GB facet structure from the micron to the nano-scale of both pure (a) and 0.3Ca-doped (b) 7° [001]-tilt bicrystals, as examined by conventional TEM. The periodically spaced dislocation cores are evident by the white-appearing structures under higher magnification in Figure 2b.

Figure 3. Atomic structures of dislocation cores for pure and 0.3Ca-doped [001]-tilt YBCO grain boundaries.

(a) Z-contrast images of [100] dislocation cores for pure 7° [001]-tilt YBCO bicrystal. The Y-Ba and Cu-O columns are indicated in a-2.
(b) Z-contrast images of [100] dislocation cores for 0.3Ca-doped 7° [001]-tilt bicrystal. The expanded, more disordered tensile region of the dislocation core is indicated in b-2.
(c) Sketch of atomic column positions at [100] dislocation cores for the pure and 0.3Ca-doped GBs. Dashed lines indicate the dislocation cores regions for pure YBCO GB and the expanded core regions for the Ca-doped GB. The gray marks three columns with anomalously intensity in the Ca-doped GB.

Figure 4. Ca segregation at the GB of the 0.3Ca-doped sample and Ca concentration variations perpendicular to and along the GB plane.

(a) Averaged Calcium $L_{2,3}$ edge spectra acquired from the 10 dislocation cores and 15 intragrain regions away from the GB
(b) Experimental data points of the Ca concentration variation in the direction perpendicular to the grain boundary. The solid curve shows the ratio $c(x)/c_0$ of the local Ca concentration $c(x)$ to the bulk value $c_0$ calculated from Eqn. (1) for y = -0.2d, and the parameters defined in the caption to Figure 6.
(c) Experimental data points (taken from the yellow contours in the top Z-contrast image) of the Ca concentration variation along the grain boundary. The solid curve in the lower panel shows the ratio $c(x)/c_0$ calculated from Eqn.1 for the same values of the parameters as those in Figures 4b and 6.

Figure 5. Oxygen EELS spectra measured on the pure and 0.3Ca-doped GBs, showing evidence for significant 528 eV oxygen pre-peak right at the GB dislocation cores.

Figure 6. Surface plot of the distribution of $Ca^{2+}$ solute ions near a periodic chain of edge dislocations in a $7°$ GB calculated from Eqn. (1), where $\varphi(x) = \varphi_0 \Sigma_n K_0[(r_0^2 + x^2 + (y - nd)^2)^{1/2}/l_D]/K_0(r_0/l_D)$, is a solution of the Thomas-Fermi equation $\nabla^2 \varphi - \varphi/l_D^2 = -4\pi q \Sigma_n f(x, y-dn)/\varepsilon$ that describes the screened potential of charged dislocation cores spaced by d along the y-axis. Here $\varphi_0 = 2q/\varepsilon$ is the amplitude of the electric potential produced by the line charge q per unit length of the core, f(r) is a function which describes the radial distribution of charge density in a region of radius ~ $r_0 \ll d$, $\int f(x,y)dxdy = 1$, and $K_0(x)$ is a modified Bessel function. The values of the parameters are the same as those used to fit the EELS data in Figure 4: $c_0 = 0.3$, $\Delta V p_0/k_B T_i = 5.6$ (taking more realistic atomic bond lengths instead of the free ionic radii can significantly increase this parameter as compared to the estimate given in the text), $r_0 = 0.5b$, $l_D = 0.55$ nm, $\Delta Z \varphi_0/k_B T_i = 26$, and $T_i \sim 300$ K, in which case $\varphi_0 \approx 0.6$ eV is comparable to $\varphi_0 \approx 1$ eV deduced from electron holography study of a $4°$ 0.2Ca-doped GB[15]. The red peaks show Ca-enrichment of tensile regions of the dislocation cores, while blue regions show Ca-depleted current channels with much weaker $T_c$ depression than in the bulk where $c_0 = 0.3$.

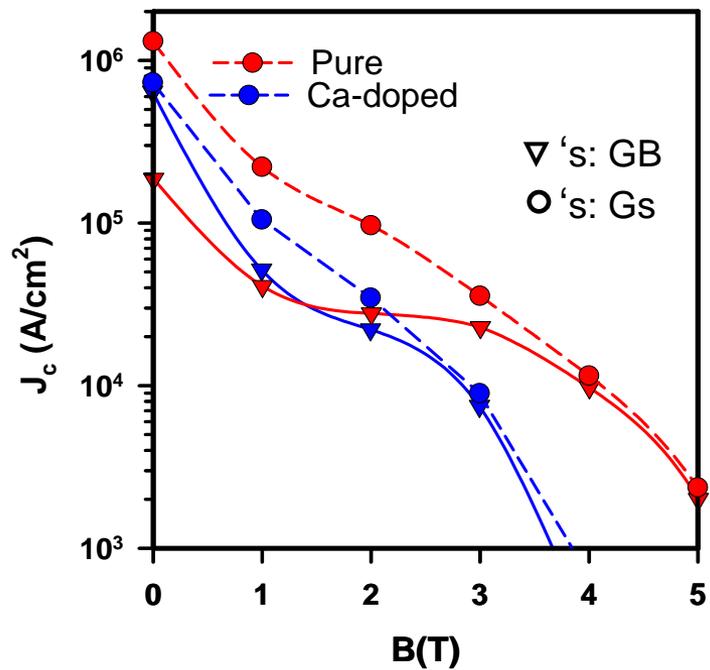
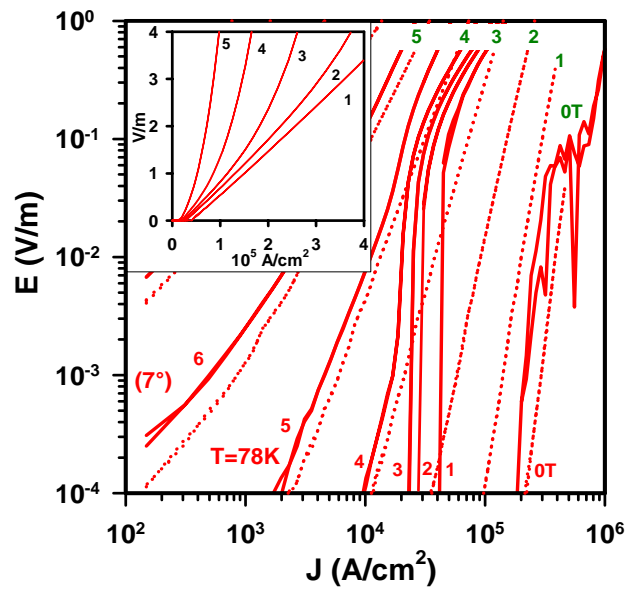
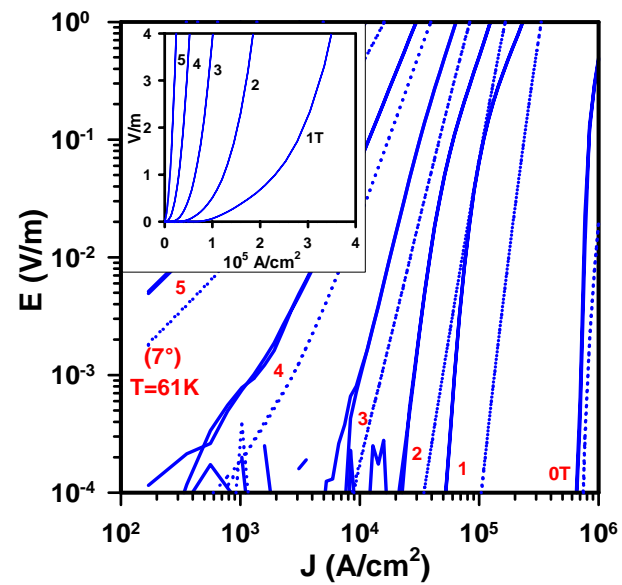

Figure 1

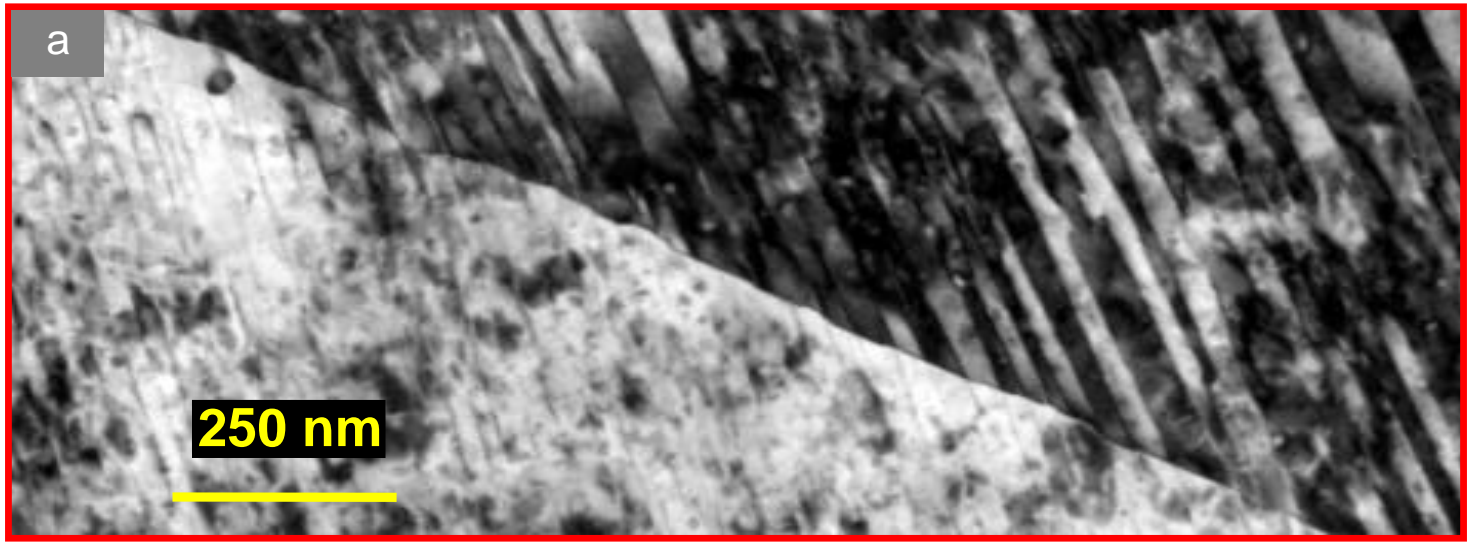
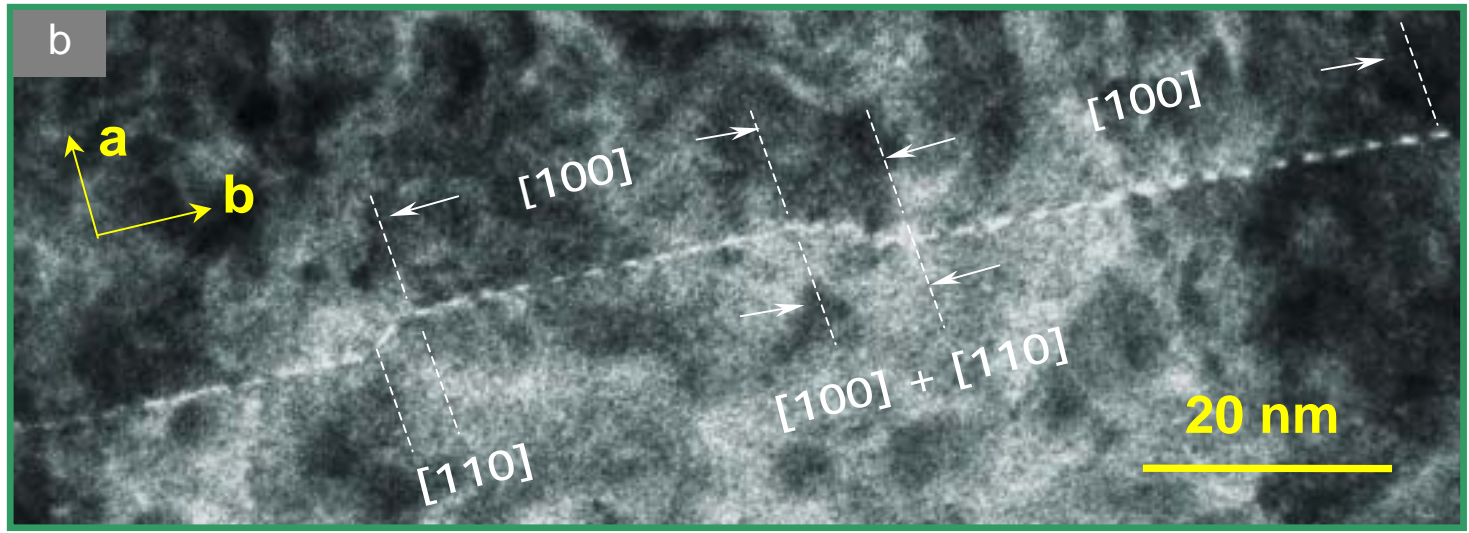

Figure 2

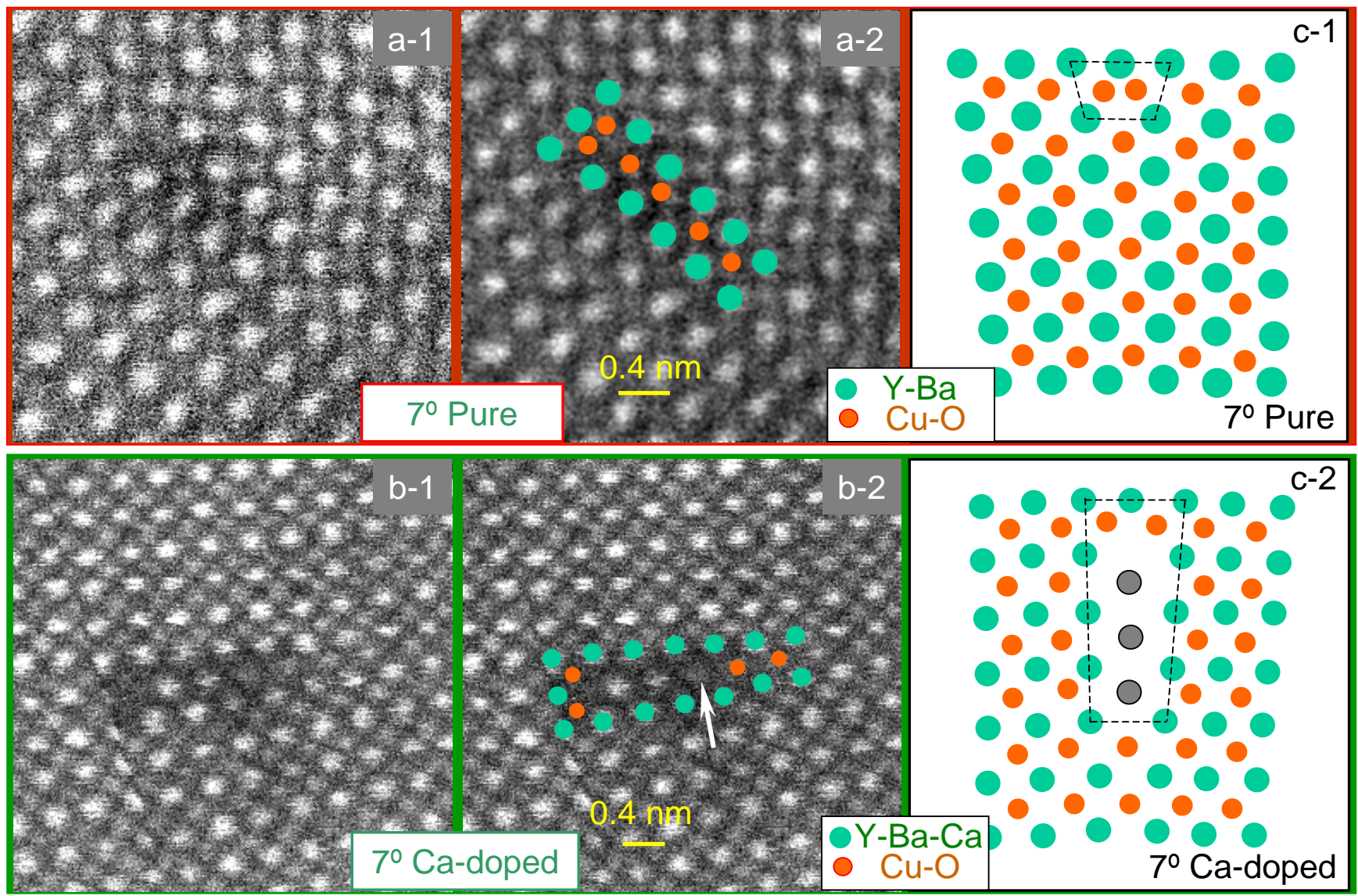

Figure 3

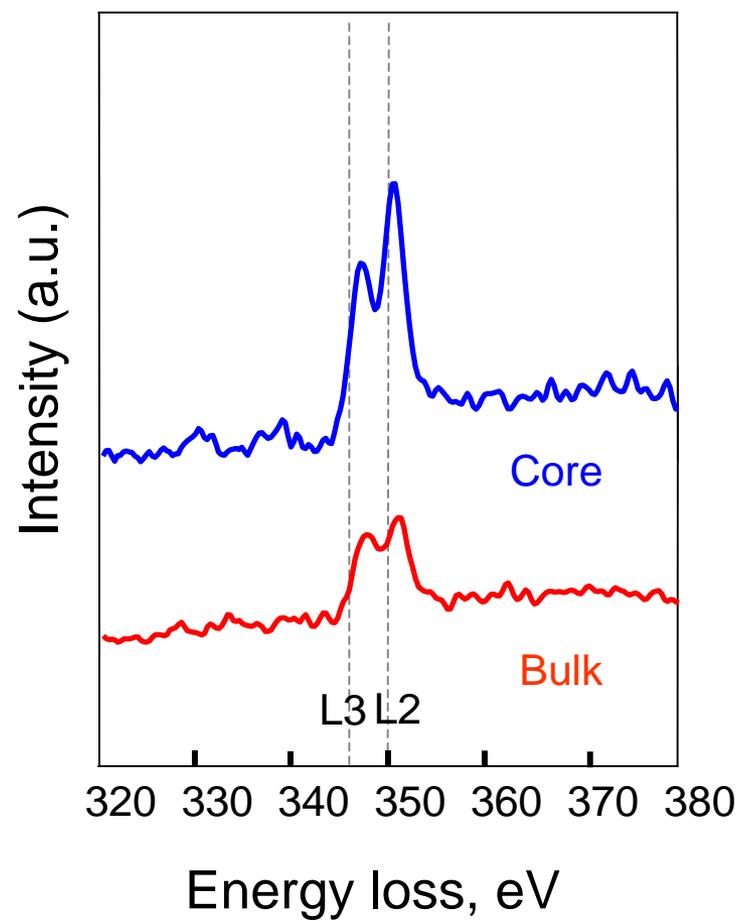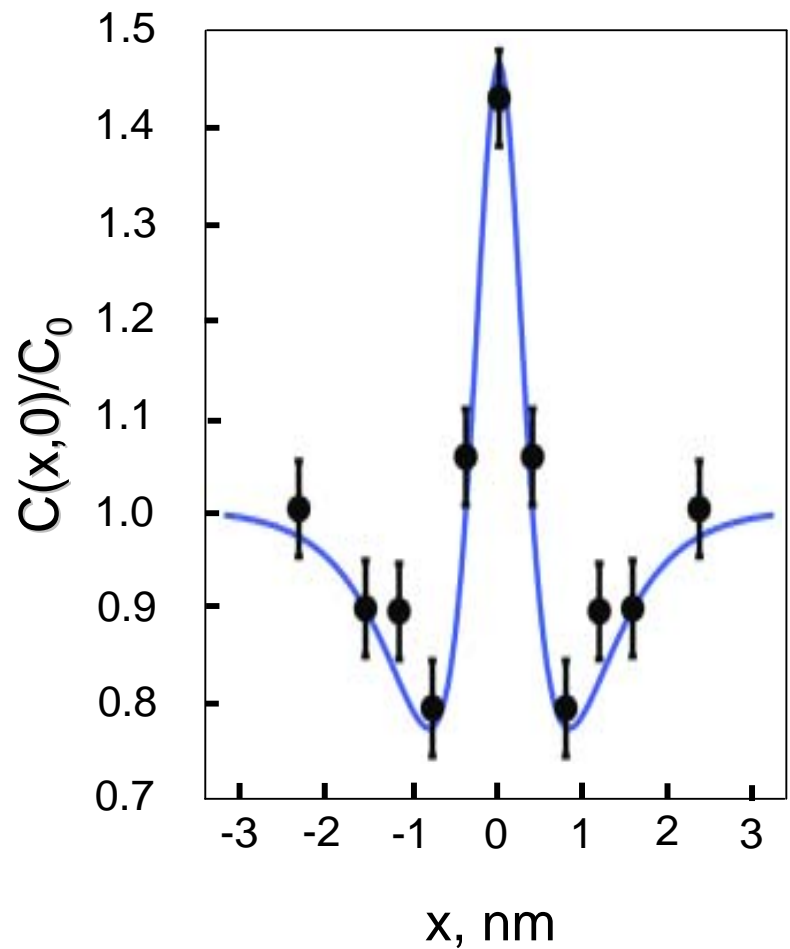

Figure 4 (a)        Figure 4 (b)

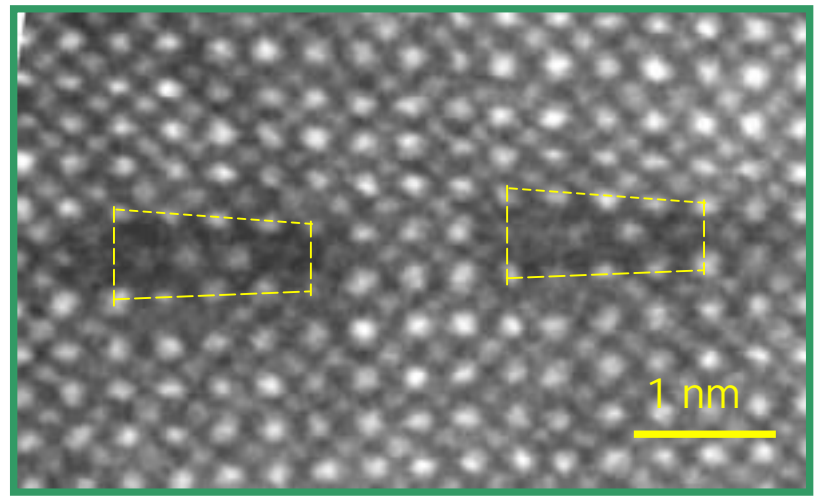
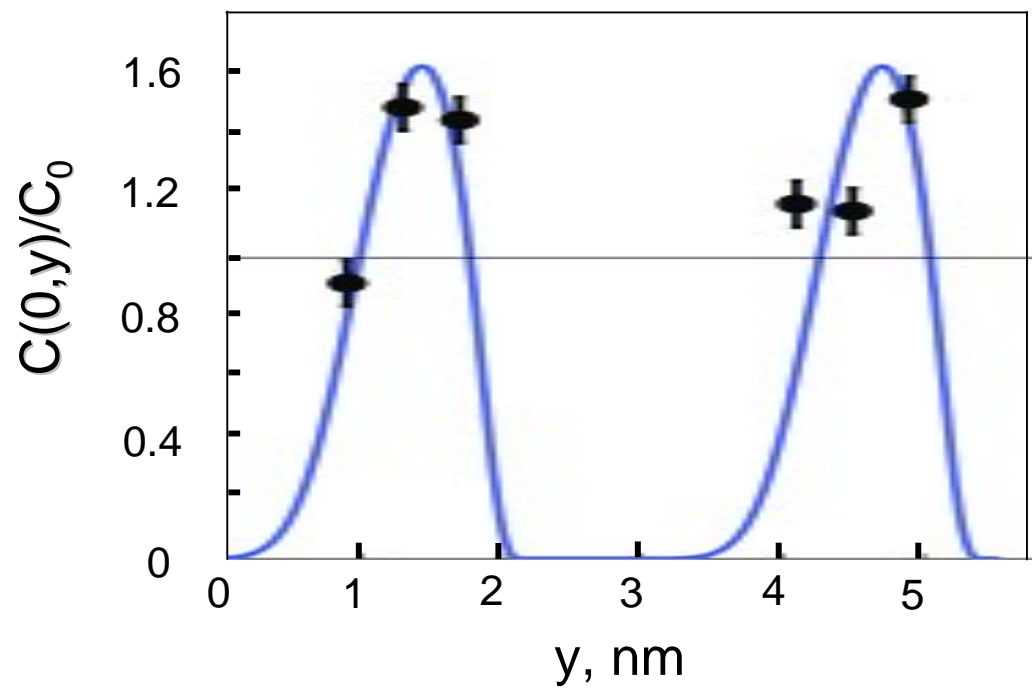

Figure 4 (c)

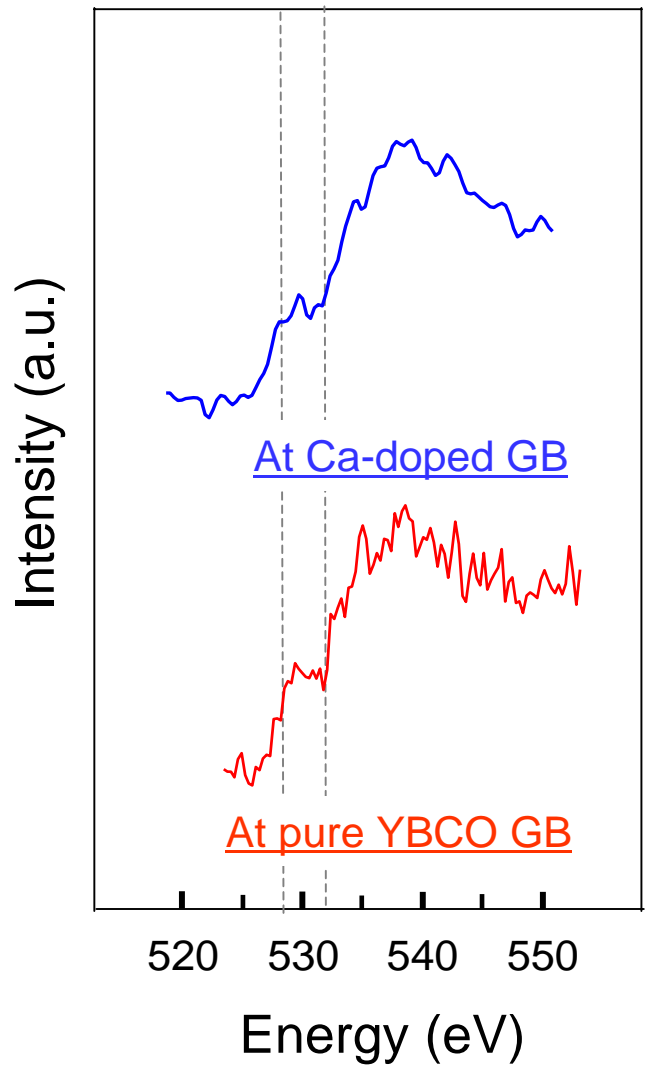

Figure 5

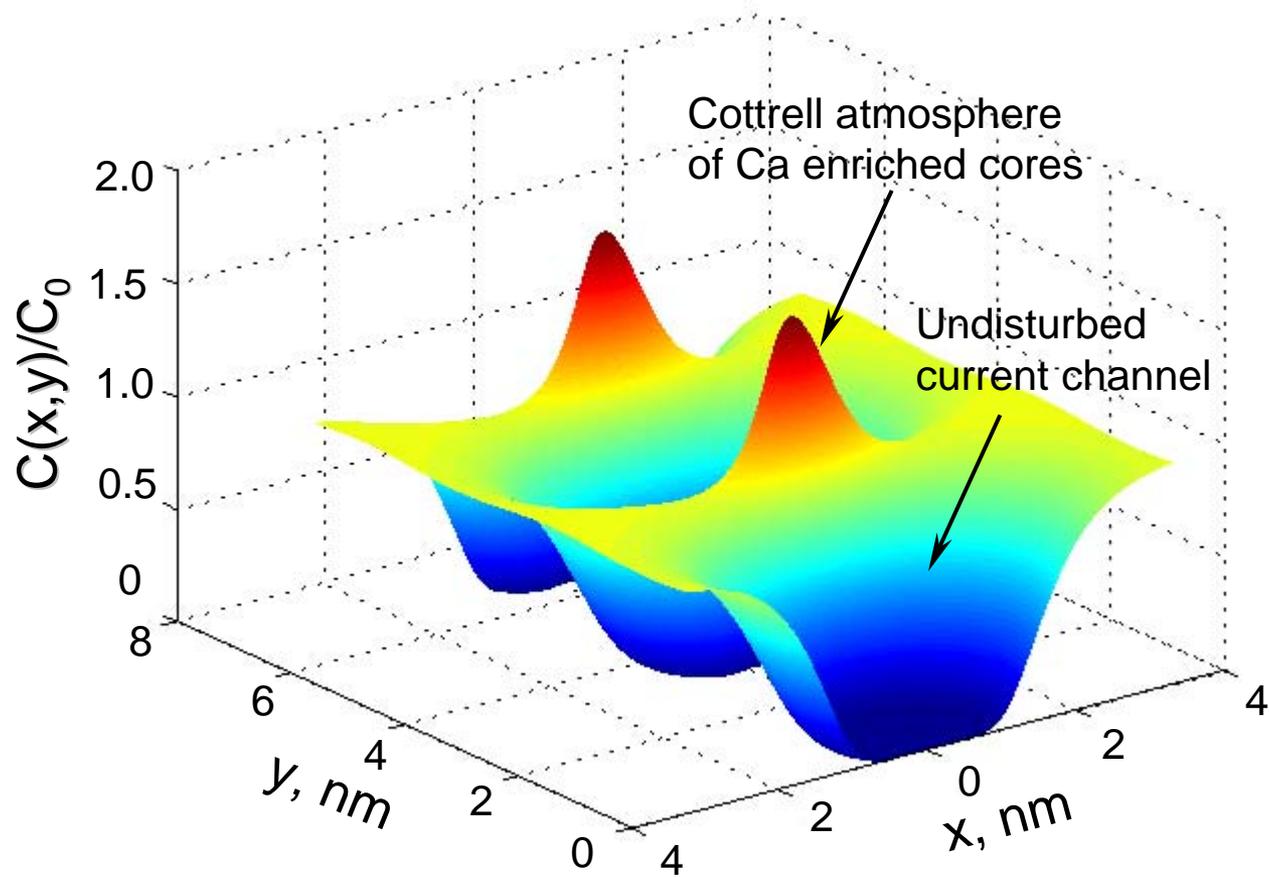

Figure 6